\documentclass[epj]{svjour}
\usepackage{epsfig} \usepackage{psfig} \usepackage{times}

\begin{document}
\title{Interactions of charmed mesons with nucleons in
the $\bar p d$ reaction \footnote{supported by DFG, RFFI and
Forschungzentrum J\"{u}lich.}}
\author{W. Cassing\inst{1}, Ye.S. Golubeva\inst{2}, L.A. Kondratyuk\inst{3}}
\institute{ Institut f\"ur Theoretische Physik, Universit\"at Giessen, \\
D-35392 Giessen, Germany
\and Institute for Nuclear Research, 60th October Anniversary Prospect
7A, \\ 117312 Moscow, Russia \and Institute of Theoretical and
Experimental Physics, B.\ Cheremushkinskaya 25, \\
117259 Moscow, Russia}
\date{Received: date / Revised version: date}

\abstract{ We study the possibility to measure the elastic $\Phi N$ (
$\Phi \equiv J/\psi, \psi(2S), \psi(3770),\chi_{2c}$) scattering cross
section in the reaction $\bar p{+}d{\to}\Phi{+}n_{sp}$ and the elastic
$D(\bar D) N$ scattering cross section in the reaction $\bar p{+}d \to
D^- D^0 p_{sp}$.  Our studies indicate that the elastic scattering
 cross sections can be determined for $\Phi$ momenta about 4-6 GeV/c
and $D/ \bar D$ momenta 2 -- 5 GeV/c by selecting events with $p_t
\geq 0.4$ GeV/c for $\Phi$'s and $p_t(p_{sp})\geq$ 0.5 GeV/c for
$D/\bar D$-meson production. \PACS{ {25.43.+t}{Antiproton-induced
reactions} \and {14.40.Lb}{Charmed mesons} \and {14.65.Dw}{Charmed
quarks} \and {13.25.Ft}{Decays of charmed mesons} }}

\authorrunning{W. Cassing et al.} \titlerunning{Interactions of
charmed mesons with nucleons in the $\bar p d$ reaction}

\maketitle

\section{Introduction}

Apart from the light flavor ($q \bar{q}$) quark physics and their
hadronic bound states the interest in hadronic states with strange
flavors ($s \bar{s}$) as well as their mutual interactions and
properties in the medium has been rising continuously in line with
the development of new experimental facilities \cite{SQM99}. In
addition to the strange quark sector also the charm quark degrees
of freedom have gained vivid interest especially in the context of
a phase transition to the quark-gluon plasma (QGP) where charmed
meson states should no longer be formed due to color screening
\cite{QM96,QM97,Satz}. However, the suppression of $J/\Psi$ mesons
in the high density phase of nucleus-nucleus collisions might also
be attributed to inelastic comover scattering (cf.
\cite{Cass99,Vogt99,Seattle98} and Refs. therein) provided that
the corresponding $J/\Psi$-hadron cross sections are in the order
of a few $mb$ \cite{Haglin}. Present theoretical estimates here
differ by more than an order of magnitude \cite{Bernd} especially
with respect to $J/\Psi$-meson scattering. Also the $J/\Psi N$
cross section is not known sufficiently well since the
photoproduction data suggest a value of 3--4 $mb$ \cite{Huefner},
while the charmonium absorption on nucleons (at high relative
momentum) in $p+A$ and $A+A$ reactions is conventionally fitted by
6--7~$mb$ \cite{Seattle98,Dimo97}. In short: the present status of
our knowledge on the interaction of charmed mesons with nucleons
-- especially at low relative momenta -- is very unsatisfying.

In this work we explore the perspectives of charmed meson -
nucleon scattering in the $\bar{p} d$ reaction because i)
antiproton annihilation will produce charmed mesons with rather
low momenta in the laboratory and ii) the momentum distribution of
the spectator nucleon in the deuteron is very well known. We
mention that related experimental studies might be carried out at
the future 'glue/charm factory' at GSI that is presently under
discussion \cite{HESR}. An alternative possibility to measure the
$J/\Psi N$ elastic scatternig cross section using pion beams in
the reaction $\pi^+ d \rightarrow J/\psi p p$ has been discussed
recently by Brodsky and Miller \cite{Brodsky}.

Our work is organized as follows: In Section 2 we will explore the
perspectives for resonance production channels of $J/\Psi, \Psi(2S),
\Psi(3770), \chi_{2c}$ and their rescattering on the spectator nucleon
including their decay to dileptons. In Section 3 nonresonance production
channels will be studied in the $\bar{p} d$ reaction with an emphasis on
$D \bar{D}$ rescattering and the option to gate on charmonium momenta by
triggering on an additional energetic photon or pion. Section 4 concludes
this study with a summary.

\section{Resonance production and rescattering in the reaction $\bar
p{+}d{\to} \Phi {+}n$}

We here examine the possibility to measure the elastic $\Phi N$ ($\Phi
\equiv J/\psi, \psi(2S), \psi(3770), \chi_{2c}$) scattering cross section
in the reaction \begin{equation} \bar p{+}d{\to}\Phi{+}n_{sp} {\to} X {+}
n_{sp}, \label{eq:Phi} \end{equation} where $\Phi$ is produced by
resonance fusion $\bar p p \to \Phi$ and $X$ the decay product of the
resonance. The main contribution to the total cross section of the
reaction (\ref{eq:Phi}) comes from the spectator term (see diagram a) in
Fig 1.), which is dominant if the momentum of the proton-spectator from
the deuteron is below 100--150 MeV/c. The amplitude corresponding to the
diagram a) of Fig.1 can be written in the deuteron rest frame as
\begin{equation} \label{eq:Ma} M_a = f(\bar p p \to \Phi \to X) \int d^3 \
{\bf r} \ e^{-i {\bf p_2} {\bf r}} \Psi_d({\bf r}), \end{equation}
where $f(\bar p p \to \Phi \to X)$ is the Breit-Wigner amplitude,
$\bf{p_2}$ the proton momentum and $\Psi_d({\bf r})$ the deuteron
wave function.

\begin{table}[htb]
\begin{center}
\caption{Charachteristics of the transitions $ \bar p p \rightarrow  \Phi
$ in vacuum and for a deuteron and carbon target. }
\vspace*{.5 cm}
\begin{tabular}{|c|c|c|c|c|}
\hline $\Phi$ & $J/\psi$ & $\Psi(2S)$ & $\Psi(3770)$  & $
\chi_{2c}(1p)$  \\ \hline $M_{\Phi}(MeV)$ & 3097 & 3686 & 3770 &
3556 \\ \hline $\Gamma_{\Phi}(MeV)$ & 0.087 & 0.277 & 23.6 & 2  \\
\hline $k_{cm}(\bar p p)(MeV/c)$ & 1232 & 1586 & 1634 & 1510 \\
\hline $T_{lab}(MeV)$ & 3250 & 5400 & 5700 & 4845 \\ \hline
$Br(\bar p p)\times 10^{-4}$ & 21.4 & 1.9 & 2 -
0.2 (?) & 1.0 \\ \hline $\sigma(\bar p p \to \Phi)(\mu b)$
& 5.15 & 0.28 & 0.7-0.07 & 0.27
\\
\hline
$Br(e^+ e^-)\times 10^{-4}$ & 600 & 85 & 0.112& - \\
\hline
$Br(\gamma J/\psi)$ & - & - & -& 0.135 \\
\hline
$S.F.(d)\times 10^{-3}$ & 1.07 & 2.6 & 221& 19.2 \\
\hline
$S.F.(C)\times 10^{-3}$ & 0.56 & 1.24 & 113& 11.2 \\
\hline
\end{tabular}
\end{center}
\end{table}

All resonances $J/\psi, \psi(2S), \psi(3770)$ and $\chi_{2c}$ are quite
narrow and the decay length for each of them produced in the reaction
(\ref{eq:Phi}) is much larger than the average distance between two
nucleons in a deuteron (see Table 1). Thus the produced meson can
rescatter elastically on the neutron spectator, transfering momenta larger
than a few hundred MeV/c. In this case the spectator term will become very
small and the contribution from the rescattering term become dominant.

Each resonance (produced in the $\bar{p} d$ reaction) has a
rather high momentum with respect to the
spectator nucleon such that we can describe the rescattering amplitude
(see diagram b) in Fig. 1) within the framework of the eikonal
approximation (cf. Ref.~\cite{Kondrat3,Shmatikov,Nikolaev}) as
\begin{equation}
 \label{eq:Mb} M_b = - f(\bar p p \to \Phi \to X) \int d^3 {\bf r} \ e^{-i
{\bf p_2} {\bf r}} \Theta(z) \Gamma({\bf b}) \Psi_d({\bf r}),
\end{equation}
where the $z$- axis is directed along the $\Phi$-momentum.
The elastic $\Phi n$ - scattering amplitude is related to the profile
function $\Gamma ({\bf b })$ by the standard expression \begin{equation}
\label{eq:fMp} f(\Phi n \to \Phi n) = \frac{1}{2\pi ik} \int d^2 {\bf b} \
e^{-i {\bf q} {\bf b}} \Gamma({\bf b}) \end{equation} with $k$ denoting
the laboratory momentum of the $\Phi$-meson and ${\bf q}$ the momentum
transfer.
\begin{figure}[htb] \vspace{-7mm}
\epsfig{file=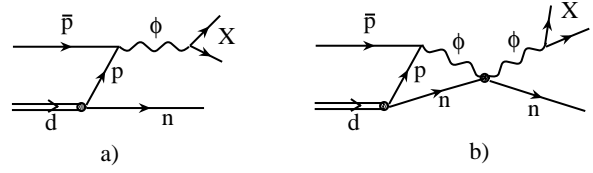,width=9.cm} \vspace{-35mm} \caption{The
diagramms for $\Phi$-meson production in the reaction
(\ref{eq:Phi}) without (a) and with $\Phi N$ rescattering (b).}
\label{fig:graphs} \end{figure}

The probability to detect the $\Phi$ through the decay channel $X$ at
solid angle $d\Omega_d$ in coincidence with the spectator of momentum $\bf
p _2$ is related to the differential cross section as \begin{equation}
\label{eq:ds/do5}
 \frac{d^5 \sigma(\bar p d \to \Phi n \to X n )}{d\Omega d^3p_2}
  = |M_a + M_b|^2 .  \label{eq:Mab} \end{equation}

In order to simulate events for the reaction (\ref{eq:Phi}) we use the
Multiple Scattering Monte Carlo (MSMC) approach. An earlier version of
this approach -- denoted as Intra-Nuclear Cascade (INC) model -- has been
applied to the analysis of $\eta$ and $\omega$ production in $\bar p A$
and $p A$ interactions in Refs.~\cite{Golub92,Golub93}. Recently this
version of the INC model has been extended to incorporate in-medium
modifications of the mesons produced~\cite{Golub96,Golub97} in
hadron-nuclear collisions.

However, the INC model is valid only for medium and heavy nuclei
and cannot be directly applied to deuterons. In order to perform
simulations of scattering events in the case of antiproton --
deuteron interactions we use the Monte-Carlo approach for the
single and double scattering terms, which are propotional to
$|M_a|^2$ and $|M_b|^2$ of Eq. (\ref{eq:Mab}), respectively. Note
that the interference between the spectator and rescattering
amplitudes is only important in a narrow region of spectator
momenta, where both contributions are of the same order of
magnitude. For a more detailed discussion of this point see, e.g.,
Refs.~\cite{Kondrat3,Shmatikov,Nikolaev,Kolybasov}.

The probability for the produced meson to rescatter on the spectator
nucleon then can be found in a standard way by \begin{equation}
 \label{eq:W} W = \frac{\sigma_{el}(\Phi n \to \Phi n)}{4\pi} r^{-2}_d
\end{equation}
with
\begin{equation}
\label{eq:Ma1}
 r^{-2}_d= \int d^3 \ {\bf r} \ r^{-2} \ |\Psi_d({\bf r})|^2.
\end{equation}
Necessary parameters for a MC simulation of rescattering are the
elastic $\Phi n$ scattering cross sections and slope parameters $b$ for
the differential cross sections $d\sigma/dt$, which are approximated by
\begin{equation} d\sigma/dt = A \exp(bt), \end{equation} where $t$ is
the momentum transfer squared. These parameters as well as the masses
of the rescattered particles determine the momentum and angular
distributions of the particles in the final state.

An important point for resonance production in $\bar{p} d$ reactions is
the rather strong dependence of the Breit-Wigner cross section $|f(\bar p
p \to \Phi \to X)|^2$ on the initial proton momentum which leads to a
rather strong suppression of the cross section on the deuteron or on
heavier targets.  The suppression factor in the maximum of the cross
section is of the order (see Ref. \cite{Farrar})  $$ S.F. \simeq \pi
\Gamma_{\Phi} m_N/(k_F M_{\Phi}),$$ where $\Gamma_\Phi, M_\Phi$ denote the
vacuum width and mass of the produced meson, whereas $k_F$ is the target
Fermi momentum. Moreover, after Fermi smearing over the target momentum
distribution the Breit-Wigner production cross section becomes much wider
and asymmetric (see e.g. Ref.~\cite{Kondrat}).

All the effects of Fermi
smearing have been taken into account in the MSMC calculations explicitly by
convoluting the resonance production cross section with the proton
momentum distribution in the deuteron (or heavier nuclei).  In a deuteron
this distribution is given by the Fourier transformed deuteron wave
function squared $|\psi_d({p})|^2$, while in medium and heavy nuclei the local
density approximation for the Fermi distribution is used.
\begin{figure}[htb]
\epsfig{file=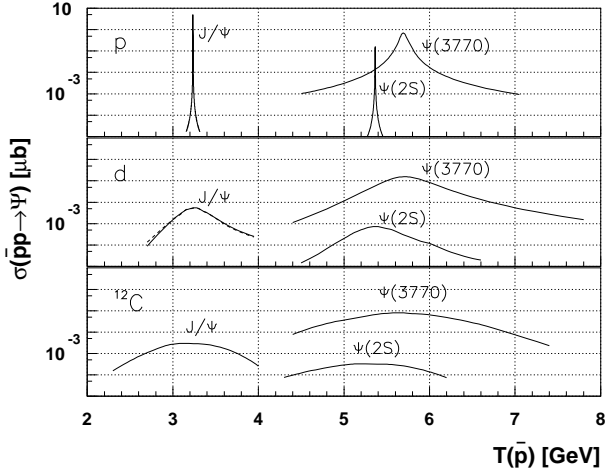,width=8cm} \caption{Fermi smearing of $\sigma(\bar
p p \to \Phi$) for $\Phi = J/\psi, \psi(2S)$ and $\psi(3770)$ in deuteron
(middle part) and carbon targets (lower part). The upper part shows the
elementary cross sections on a hydrogen target.} \label{fig:fermi}
\end{figure}

In Fig. \ref{fig:fermi} we present the cross sections $\sigma(\bar p p \to
J/\psi)$, $\sigma(\bar p p \to \psi(2S))$ and $\sigma(\bar p p \to
\psi(3770))$ as a function of the antiproton kinetic energy for a hydrogen
target (upper part), for a deuteron target (middle part) and carbon target
(lower part). In Fig. 2 two excitation curves for $J/\psi$ production on
the deuteron are shown to demonstrate the dependence on the choice of the
deuteron wave function: the dashed line corresponds to the Paris model
\cite{Lacomb}, while the solid curve is calculated using the H\"ulthen
wave function \begin{equation} \label{eq:Hulthen} \Psi_d (r) = N
\left(e^{- \alpha r} - {e^{-\beta r}} \right)/r, \end{equation} where N is
the normalization factor, $ \alpha = \sqrt{m \epsilon} \simeq ~46.5 ~MeV/c
$, $ \beta~\simeq ~{5.2 \alpha} $.  One can see that this dependence is
quite small since the two results practically overlap.

\begin{figure}[t] \epsfig{file=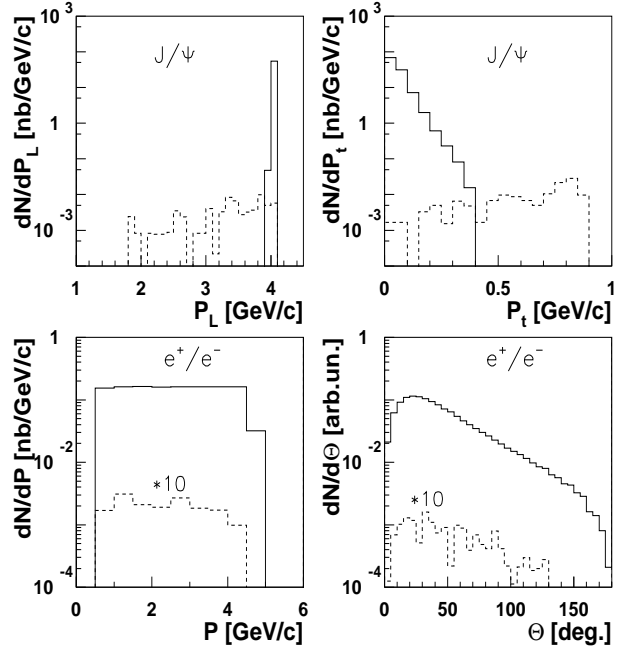,width=8cm,height=8.6cm}
\caption{Distributions of $J/\psi$ mesons from the reaction
(\protect\ref{eq:J/psi}) in longitudinal (upper left) and transverse
(upper right) momentum for $T_{lab}$ = 3.25 GeV. The distribution of the
decay products $e^+/ e^-$ in the total momentum is shown additionally
(lower left) as well as in the polar angle $\theta$ (lower right). The
solid and dashed curves describe the contributions of the spectator and
rescattering terms, respectively. The $J/\psi n$ elastic scattering cross
section was assumed to be 1 mb and the differential slope parameter was
taken as $b \simeq$1 (GeV/c)$^{-2}$.} \label{fig:f2} \end{figure}

Let us now consider the reaction \begin{equation}
 \label{eq:J/psi} \bar p d \to J/\psi n \to e^+ e^- n \end{equation}
In Fig. \ref{fig:f2} we show the distributions of $J/\psi$ in
longitudinal (upper left) and transverse (upper right) momentum as
well as the momentum distribution of the decay products $e^+/ e^-$
(lower left) and their distribution in the polar angle $\theta$
(lower right) for $T_{lab}$ = 3.25 GeV. The solid and dashed
curves describe the contributions of the spectator and
rescattering terms, respectively. The $J/\psi \ n$ elastic
scattering cross section here was assumed to be 1 mb while the
slope parameter was taken as $b \simeq$1 (GeV/c)$^{-2}$, which is
almost the same as in the reaction $\gamma p \to J/\psi p$
\cite{Gittelman} at roughly the same $s \simeq$ 20 GeV$^2$.

It is seen that after rescattering on the neutron the $J/\psi$ looses
longitudinal momentum (upper left part), but gains transverse momentum
(upper right). The momentum spectrum of the leptons from $J/\psi$ decay is
completely flat (lower left part), while their angular distribution (lower
right) has a broad maximum around 25$^o$.

It is clear that performing a cut for $p_t(J/\psi)\geq 400$ MeV/c or
$p_L(J/\psi) \leq 3.6$ GeV/c one can determine essentially the overall
magnitude of the $J/\psi n$ elastic cross section. This is demonstrated in
Fig. 4 where we show the number of events as a function of the cut in the
$J/\psi$ transverse momentum. The dotted and dashed curves describe the
tails of the spectator momentum distribution for the Paris and H\"ulthen
models of the deuteron wave function, respectively. For $p_t(J/\psi)\geq$
400 MeV/c both curves drop far below the solid histograms which describe
the contribution of the $J/\psi n$ rescattering term. The relative number
of events with rescattering is propotional to $\sigma_{el}(J/\psi n)$ and
for the cut $p_t(J/\psi)\geq 400$ MeV/c is about 0.1 $\%$ using
$\sigma_{el}(J/\psi n)$ = 1 mb.

\begin{figure}[htb] \centerline{\epsfig{file=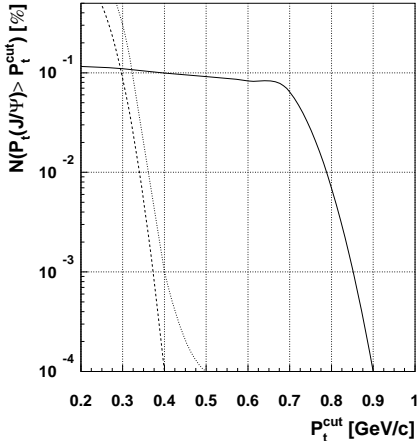,width=5.5cm}}
 \caption{The number of events for reaction (\protect\ref{eq:J/psi}) as a
function of the $p_t$ cut for $p_t(J/\psi)\geq p_t^{cut}$. The dotted and
dashed curves describe the contributions of the spectator term for the
Paris and H\"ulthen models of the deuteron wave function, respectively.
The solid curve describes the rescattering term for $\sigma_{el}(J/\psi
n)$ = 1 mb and a slope parameter for the differential cross section $b$ =
1 (GeV/c)$^{-2}$.} \label{fig:Jcut} \end{figure}

In a similar way we can evaluate the resonance production and rescattering
of particles from the $\Phi$ family as $\psi(2S)$,$\psi(3770)$ and
$\chi_{2c}(1p)$ which can be detected via their decays $l^+ l^-$, $D \bar
D$ or $\gamma J/\psi$, respectively. The suppression factors due to Fermi
smearing of the cross sections ${\bar p} p \to \Phi$ in deuteron and
carbon targets for all particles are shown in Table 1.

The case of the $\psi(2S)$ looks much less promising than the case for
$J/\psi$, because its signal in the $l^+ l^-$ channel will be smaller by
almost two orders of magnitude. A much stronger signal of the $\psi(2S)$
will be in the channel $J/\psi \pi^+ \pi^-$, but in this case a
measurement of the $\psi(2S)$ transverse momentum is difficult.

The signal of the $\chi_{2c}(1p)$ in the channel $\gamma J/\psi$ will be
almost the same as the signal of the $J/\psi$ in the dilepton channel.
Despite of its smaller production cross section in ${\bar p} p$ the
$\chi_{2c}(1p)$ is not so strongly suppressed by Fermi smearing as in case
of the $J/\psi$ (cf. Table 1). Therefore, if the photon from the
$\chi_{2c}(1p)$ decay can be detected in coincidence with the dileptons from
the $J/\psi$ decay, a measurement of the $\chi_{2c}(1p) n$ elastic
cross section can also be performed by selecting events with
$p_t(\chi)\geq 400$ MeV/c.

It is, furthermore, interesting to consider the reaction \begin{equation}
 \label{eq:3770} \bar p d \to \psi(3770) N \to D \bar D N .
\end{equation} Up to now the resonance $\psi(3770)$ has been seen only in
the charge neutral mode in $e^+ e^-$- collisions. Therefore, its isospin
is not yet known and a detection of $D \bar D$ production in $\bar p n$
would unambiguously fix its isospin as 1. If the $D \bar D$ pair will be
produced only on a proton, then its isospin will be 0.

The coupling of the $\psi(3770)$ to the $p \bar p$ channel is not known.
Having in mind that close to the $\psi(3770)$ the resonances
$\chi_{2c}(1p)$ and $\psi(2S)$ have a branching ratio $BR(\bar p p) \simeq
2.10^{-4} $ (despite of their full widths being different by almost an
order of magnitude), it seems to be justified to assume $BR(\psi(3770) \to
\bar p p) \simeq 2.10^{-4}- 2.10^{-5} $. Then the cross section of its
production on a nucleon in the deuteron will be about 15--150 nb.

\begin{figure}[t]
\epsfig{file=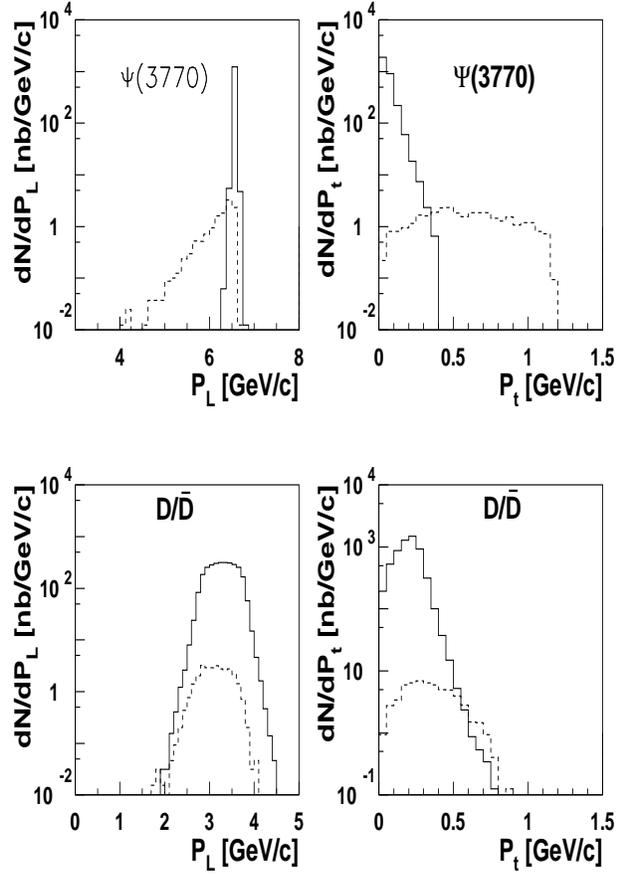,width=8cm,height=11.6cm}
 \caption{Momentum distributions in $p_L$ (left) and $p_t$ (right) for
$\psi(3770)$ (upper) and $D/ \bar D$ (lower) from the reaction
(\protect\ref{eq:3770}). The solid and dashed curves describe the
contributions of the spectator and rescattering terms, respectively, with
$\sigma(\bar p p \to \psi(3770 ))$ =150 nb, $\sigma(\psi(3770) N \to
\psi(3770) N)$ =10 mb and a slope parameter $b(\psi(3770)N $= 2
(GeV/c)$^{-2}$.} \label{fig:f4} \end{figure}

In Fig.\ref{fig:f4} we present the simulated momentum distributions of
$\psi(3770)$ and $D/ \bar D$ mesons for $\sigma(\bar p p \to \psi(3770 ))$
=150 nb, $\sigma(\psi(3770) N \to \psi(3770) N)$ =10 mb and a slope
parameter $b(\psi(3770)N) $= 2 (GeV/c)$^{-2}$. The solid and dashed curves
describe the contributions of the spectator and rescattering terms,
respectively. The $\psi (3770)$ distributions in the longitudinal and
transverse momentum are shown in the upper left and upper right part,
respectively. The rescattering term dominates for $p_L \leq$ 6 GeV/c and
$p_t \geq$0.4 GeV/c. The $p_l$ and $p_t$ spectra of $D/\bar D$ mesons are
presented in the lower left and lower right part of the figure. The change
of the longitudinal and transverse momentum
spectra for $\psi(3770)$ due to rescattering has
also some influence on the $p_l$ and $p_t$ spectra of the $D/\bar D$
mesons especially for the lowest $p_l$ and the highest $p_t$. However, the
signal of rescattering is much cleaner in the $p_t$ distribution of the
$\psi(3770)$ for $p_t \geq$0.4 GeV/c: practically there is no background from
the spectator mechanism and the relative number of the events is 0.73
$\%$ in case of the cross sections quoted above.

\section{Nonresonance production and rescattering of $J/\psi$ and $D/\bar
D$}

Since there is a quite large suppression in the resonance production for
narrow quarkonium states on nuclear targets, their nonresonance production
(where a suppression factor is absent) might well be comparable to the
resonant formation. In this Section we thus consider the rescattering of
$J/\psi$ and $D/\bar D$ in the deuteron where the mesons are produced in
the nonresonant reactions

\begin{equation}
 \label{eq:Jpi} \bar p d \to J/\psi \ \pi^- p \end{equation}
\begin{equation}
 \label{eq:Jgamma} \bar p d \to J/\psi \ \gamma n \end{equation}
\begin{equation}
 \label{eq:DDbar} \bar p d  \to D \bar D N.  \end{equation}

\begin{figure}[t]
\epsfig{file=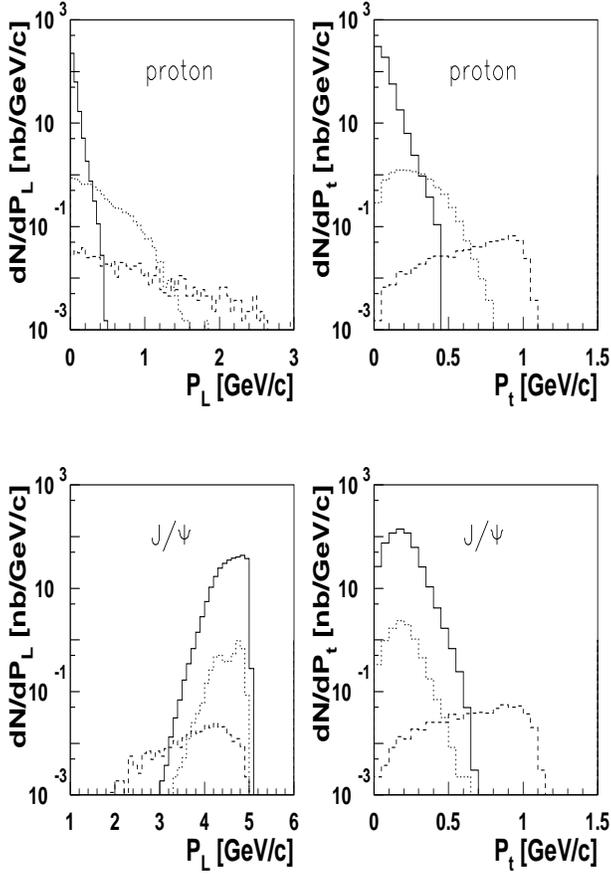,width=8cm,height=11.6cm}
\caption{Longitudinal and transverse momentum distributions of
$J/\psi$ (lower part) and the proton spectator (upper part) from
the reaction (\protect\ref{eq:Jpi}) at $T_{lab}$ =5 GeV. The solid
curves are the contributions from the spectator term. The
contributions from pion and $J/\psi$ rescattering on the proton
spectator are shown by dotted and dashed curves, respectively. The
parameters of the $J/\psi N$ scattering amplitude were taken the
same as in Fig. 3.} \label{fig:f5} \end{figure}

The cross section of the reaction (\ref{eq:DDbar}) has been
calculated in Ref. \cite{Kaidalov} within the framework of the
Quark-Gluon String Model. It grows from threshold up to $p_{lab}
\simeq 7.5$ GeV/c, where it reaches its maximal value of about 40
nb, and then decreases rather fast with energy as
$$s^{2[\alpha(0)_{\Sigma_c}-1]}, $$ where $\alpha(0)_{\Sigma_c}$
is the intercept of the $\Sigma_c$ Regge trajectory
$$\alpha(t)_{\Sigma_c} \simeq -1.82 + 0.5 t .$$ The cross sections
of the reactions (\ref{eq:Jpi}) and (\ref{eq:Jgamma}) in the
continuum are not known. We assume that $\sigma(\bar p d \to
J/\psi \ \pi^- p)$ is comparable with $\sigma( \bar p d \to D
\bar{D} N)$ and take for our estimates 30 nb at $T_{lab}$ = 5 GeV.

\begin{figure}[t]
\epsfig{file=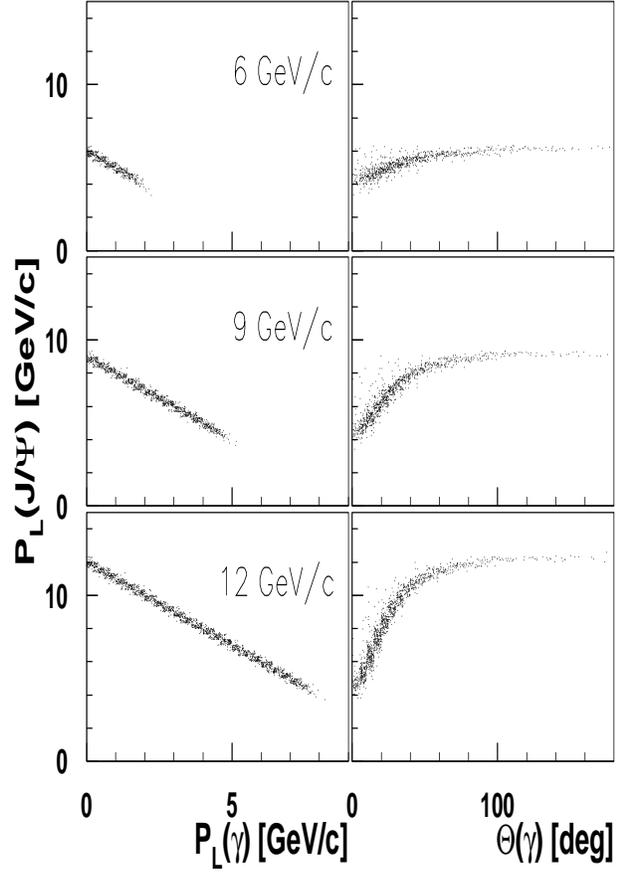,width=8cm,height=11.6cm} \caption{
Correlation between the longitudinal momenta of $J/\psi$'s and
photons at $p_{lab}$ = 6, 9 and 12 GeV/c (l.h.s.) and the angular
distributions of photons at the same momenta in the reaction
(\protect\ref{eq:Jgamma}).} \label{fig:f6} \end{figure}

The cross section for the reaction $\bar p p \to J/\psi \ \gamma $ can be
estimated at $T_{lab}$ = 4.845 GeV where it is determined through the
resonance $\chi_{2c}(1p)$ (cf. Table 1); it is about 30-40 nb. Of course,
in the continuum it is expected to be smaller.

The longitudinal and transverse momentum spectra of $J/\psi$ and the
proton spectator from the reaction (\ref{eq:Jpi}) at $T_{lab}$ =5 GeV are
shown in Fig. \ref{fig:f5}. The solid curves describe the contributions from
the spectator term, the dottted and dashed histograms describe the
contributions from pion and $J/\psi$ rescattering on the proton spectator.
The parameters of the $J/\psi N$ scattering amplitude were taken the same
as in Fig. 3. The initial antiproton energy was chosen to be above the
$J/\psi \pi$ threshold, but slightly below the $J/\psi \pi \pi$ threshold.
Despite the kinematics of the reaction (\ref{eq:Jpi}) is less resctrictive
and the $J/\psi$ has a broader distribution in transverse momentum, it is
still possible -- using the cut $p_t(J/\psi \geq$ 0.7 GeV/c -- to obtain a
rather clean signal from the $J/\psi N$ rescattering events.

Similar considerations can be made for the reaction (\ref{eq:Jgamma}). In
this case it is possible to use the photon momentum as an additional
trigger for $J/\psi n$ rescattering events, which can be used to fix the
momentum of the $J/\psi$. The correlation between the longitudinal momenta
of $J/\psi$'s and photons is presented for different momenta of the
antiproton in Fig. \ref{fig:f6} (l.h.s.). As compared to the reaction
(\ref{eq:Jpi}) one looses about 2 orders of magnitude in event rate due to
the electromagnetic vertex, but the descrimination from background events
is rather clean. The angular distributions of photons in this reaction are
also presented in Fig. 7 (r.h.s.) and show that the photons are
predominantly emitted to forward angles.

\begin{figure}[t]
\epsfig{file=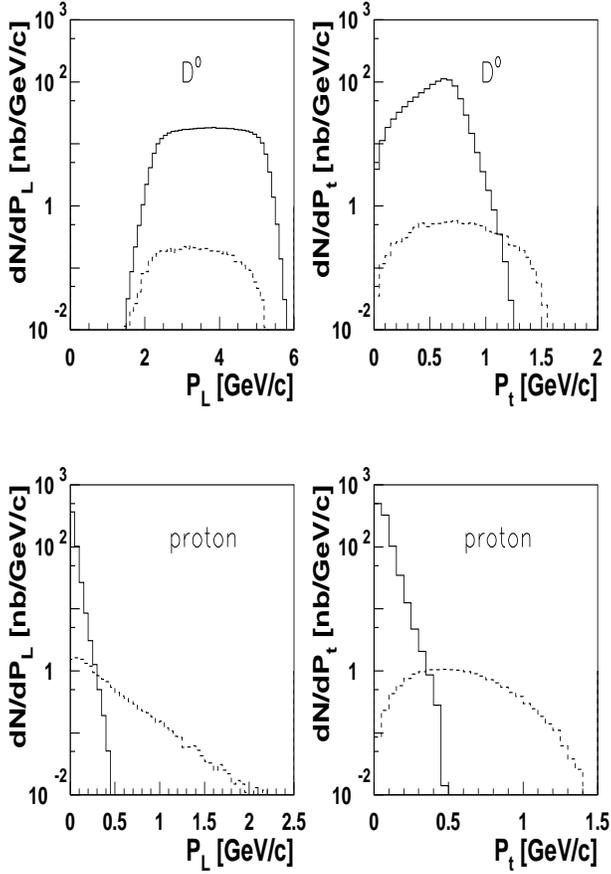,width=8cm,height=11.6cm}
 \caption{Longitudinal and transverse momentum distributions of $D^0$-
mesons (upper part) and proton spectators (lower part) from the reaction
(\protect\ref{eq:DDbar}) at $p_{lab}$ =7.5 GeV/c. The solid curves are the
contributions from the spectator term while the contributions from $D^0 N$
rescattering on the proton spectator are shown by the dashed histograms.
The elastic $D^0 p$ scattering cross section and the slope parameter $b$
were assumed to be 10 mb and 2 (GeV/c)$^{-2}$, respectively.}
\label{fig:f7} \end{figure}

Let us now discuss the possibility to measure the $DN$ and $\bar D N$
elastic scattering cross sections in the reaction (\ref{eq:DDbar}). In
Fig. 8 we show the longitudinal and transverse momentum distributions of
$D^0$- mesons (upper part) and proton spectators (lower part) at $p_{lab}$
=7.5 GeV/c. The solid curves are the contributions from the spectator term
while the contributions from $D^0 N$ rescattering on the proton spectator
are shown by the dashed histograms. The elastic $D^0 p$ scattering cross
section and the slope parameter $b$ were assumed to be 10 mb and 2
(GeV/c)$^{-2}$, respectively. In the case of $\bar D N$ scattering we have
used $\sigma_{el}$ = 5 mb and the same slope parameter $b$. In this case
the contribution of the $D^- p$ scattering term is smaller by a factor of
2. It is seen that the spectator term dominates for $p_{sp} \leq $ 400
MeV/c. However, at $p_{sp} \geq $500 MeV/c the main contribution to the
spectrum comes form the rescattering term. The individual
contributions from $D^0 p$ and $D^- p$ rescattering can be separated using
the correlation between the azimuthal angles of the two scattering planes
$p_{lab}$-- $p_{D/\bar D}$ and $p_{lab}$--$p_{sp}$.

\begin{figure}[t]
\epsfig{file=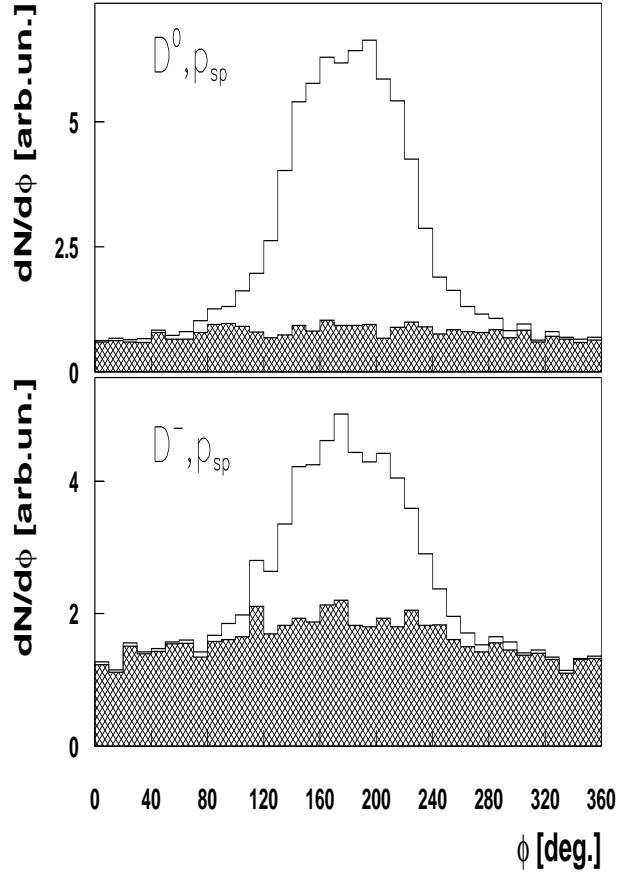,width=8cm,height=11.6cm}
\caption{ The distributions in the azimuthal angle $\phi$ between
the planes $p_{lab}$-- $p_{D^0}$ and $p_{lab}$--$p_{sp}$ (upper
part) and between the planes $p_{lab}$-- $p_{D^-}$ and
$p_{lab}$--$p_{sp}$ (lower part) for $p_t(p_{sp}) \geq$ 500 MeV/c
in the reaction $\bar{p} d \rightarrow D\bar{D}N$ at $p_{lab}=$
7.5 GeV/c.} \label{fig:f8} \end{figure}

This correlation is presented in Fig. \ref{fig:f8} for $p_t(p_{sp}) \geq$
500 MeV/c where we show the distribution of events in the azimuthal angle
between the planes $p_{lab}$-- $p_{D^0}$ and $p_{lab}$--$p_{sp}$ in the
upper part and between the planes $p_{lab}$-- $p_{D^-}$ and
$p_{lab}$--$p_{sp}$ in the lower part. The shaded area describes the
background from $D^- p$ (upper part) and $D^0 p$ rescattering (lower
part).

\section{Summary} In this study we have explored the perspectives of
measuring the elastic cross section of charmed mesons with the spectator
nucleon in resonant and nonresonant $\bar{p} d$ reactions. Our analysis
within the MSMC approach indicates that the elastic scattering cross
sections can be determined for $\Phi (\equiv J/\psi,$ $\psi(2S)$,
$\psi(3770),$ $ \chi_{2c})$ momenta about 4-6 GeV/c and $D/ \bar D$ momenta
of 2 -- 5 GeV/c by selecting events with $p_t \geq 0.4$ GeV/c for $\Phi$'s
and $p_t(p_{sp})\geq$ 0.5 GeV/c for $D/\bar D$-meson production.

We mention that the inelastic cross sections of charmed mesons may
be studied in $\bar{p}A$ reactions as analysed in Ref.
\cite{Alex99}. This opens interesting perspectives for a future
high energy antiproton storage ring \cite{HESR}.

\section*{Acknowledgements} We are grateful to A. B. Kaidalov, W. K\"uhn
and A. Sibirtsev for helpful discussions and valuable suggestions. This
work was supported by DFG, RFFI and INTAS grant No. 96-0597.

\end{document}